\documentclass[final,3p,times,sort&compress,twocolumn]{elsarticle}
\usepackage{graphicx}
\usepackage{array,booktabs,hyperref,subfig}
\usepackage[thinspace,thinqspace,squaren,textstyle]{SIunits}
\usepackage{amsmath,amssymb,braket,feynmp,slashed}
\usepackage{amsfonts,dsfont,mathrsfs}
\journal{arXiv.org}
\begin{document}%

\begin{frontmatter}%
\title{t3evol -- Numerical Solution of Twist-three Evolution Equations}
\author{B.\ M.\ Pirnay}
\address{
  Institut f{\"u}r Theoretische Physik, Universit{\"a}t Regensburg, D-93040 Regensburg, Germany
}
\date{\today}
\begin{abstract}%
A program dedicated to the numerical solution of the evolution equations for twist-three multiparton correlation functions is presented.
The solutions are obtained by direct integration on a discretized momentum fraction grid.
Both flavor nonsinglet and flavor singlet evolution (in both $C$-parity sectors) can be addressed.
Physical applications include single spin asymmetries and the subleading twist contribution to the polarized structure function $g_2$.
An arbitrary input is accepted for the initial distributions.
\section*{Program Summary}
\noindent
{\em Title of program\/}: {\tt t3evol}\\
{\em Version\/}: 1.0 \\
{\em Catalogue identifier\/}: \\
{\em Program obtainable from\/}: {\tt http://arxiv.org/archive/hep-ph} or its mirrors by downloading the source of this document.\\
{\em Distribution format\/}: tar.gz \\
{\em E-mail\/}: {\tt bjoern.pirnay@physik.uni-r.de} \\
{\em Licensing provisions\/}: GNU General Public License. \\
{\em Computers\/}: all. \\
{\em Operating systems\/}: all. \\
{\em Program language\/}: {C++}. \\
{\em Other programs called\/}: none. \\
{\em External files needed\/}: none. \\
{\em No.\ of lines in distributed program, including test data, etc.\/}: 7018\\
{\em No.\ of bytes in distributed program, including test data, etc.\/}: 269445\\
{\em Keywords\/}: Twist-three correlations, evolution equations.\\
{\em Nature of the physical problem\/}: Solving the evolution equations for the twist-three antiquark-gluon-quark and triple-gluon correlation functions with leading order kernels.\\
{\em Method of solution\/}: Series expansion of the formal solution, each term in this series typically requires several integrations, which are performed numerically on a discretized grid.\\
{\em Restrictions on complexity of the problem\/}: Momentum grids of size $N_p>300$ (see text) are not feasible as they will result in a very long computation time.\\
{\em Memory required to execute\/}: Strongly dependent on the number of grid points; for $N_p=101$ (see text) up to $15\,\text{MB}$. \\
{\em Typical running time\/}: Strongly dependent on the number of grid points; for $N_p=101$ (see text) on a PC with a 3.40 GHz processor the time is $<\unit{15}{\second}$.
\end{abstract}%

\end{frontmatter}%

\section{Introduction}

By virtue of factorization theorems, observables in hadronic processes are typically represented as a convolution of a short distance partonic part with long distance distributions.
The former can be calculated reliably in perturbative Quantum Chromodynamics (QCD), while the latter do not admit such treatment.
Both parts depend on a separation (factorization) scale $\mu$, and although the long distance parton distributions are non-perturbative, their scale dependence can be calculated perturbatively.
The corresponding differential equations, dubbed evolution equations, are indispensable in high energy phenomenology.

A prime example for this concept are the Dokshitzer-Gribov-Lipatov-Altarelli-Parisi (DGLAP) equations for the usual collinear parton distribution functions (PDFs).
Typically one invents a model or a suitable parametrization for the PDFs at some low reference scale and evolves them to the relevant scales at which experimental data is available.
One may fit the parameters such that the data is described sufficiently well.
This approach has become customary for the DGLAP setup, where the evolution is governed by the anomalous dimensions of leading twist operators.
Many programs are available to perform the evolution numerically, see e.g.~\cite{Vogt:2004ns,Cafarella:2003jr,Salam:2008qg,Botje:2010ay}.

However, large single spin asymmetries in polarized reactions have been consistently observed in a large variety of experiments, cf.~\cite{Liang:2000gz} for an overview, which can not be explained by the leading twist PDFs, see~\cite{Liang:2000gz,Anselmino:1994gn,Barone:2001sp}.
One possible explanation of the asymmetries, that goes beyond the simple PDF description, relies on large parton correlations inside the nucleon.
Factorization theorems for this framework have been worked out, cf.~\cite{Efremov:1981sh,Qiu:1991wg}.

These correlations are defined via matrix elements of a set of twist-three operators and the corresponding evolution equations have been derived in leading order of the strong coupling, see~\cite{Bukhvostov:1984as,Bukhvostov:1985rn,Balitsky:1987bk,Braun:2009mi,Ma:2012xn}.
The situation is more complicated compared to the twist-two sector, since in general one has to deal with functions of two (or more) momentum fractions.
There seems to be a tendency to consider only certain values of the momentum fractions, like the soft-gluon-pole limit~\cite{Kang:2008ey,Vogelsang:2009pj,Schafer:2012ra,Kang:2012em}.
However the corresponding evolution equations are incomplete.
For a closed set of equations all ingredients (evolution kernels and correlation functions) have to be given as functions of arbitrary momenta.
Technically, the solution of the evolution equations gets more involved compared to e.g.\ the DGLAP-scenario.
On the conceptual level, the invention of a parametrization or a model for all possible momenta is rather in its infancy.
First steps have been made in~\cite{Braun:2011aw}.
Consequently, many existing data analyses either do not include evolution at all or revert to a simplified leading twist evolution.
Either of these treatments lack theoretical justification.
The general case will be addressed here.

In this work a computer program, named {\tt t3evol}, is presented that may either be used directly in the data analysis or serve as a guideline toward a correct treatment of the QCD evolution.
The presentation is organized as follows.
Sec.~\ref{sec:formulation} recapitulates the structure of the twist-three evolution and introduces necessary notations.
The method of solution is discussed in Sec.~\ref{sec:method}.
The actual program, its configuration and certain possible modifications to it are described in Sec.~\ref{sec:description}.
Sec.~\ref{sec:output} is devoted to the explanation of the output generated by {\tt t3evol}.
Conclusions are drawn in Sec.~\ref{sec:conclusion}.

\section{Formulation of the problem\label{sec:formulation}}

The evolution equations for the twist-three correlation functions have been derived in~\cite{Bukhvostov:1984as,Bukhvostov:1985rn,Balitsky:1987bk} and reexamined in~\cite{Kang:2008ey,Braun:2009mi,Schafer:2012ra,Ma:2012xn,Kang:2012em} in different operator bases.
In one of the choices one considers operators
\begin{align}
 S^\pm_{q,\sigma}(z) &=  \bar q(z_1) t^a\bigl(i\tilde F_{\sigma+}^a(z_2)\pm F_{\sigma+}^a(z_2)\gamma_5\bigr)\gamma_+ q(z_3)\,,\notag\\
 F^\pm(z) &= 2g C^{abc}_\pm s^\sigma_T F_+^{\,\,\nu,a}(z_1)F_{+\sigma}^{b}(z_2)F_{+\nu}^{c}(z_3)\,.
\end{align}
Here, the subscript ``$+$'' stands for the contraction of a Lorentz index with a light-like vector $n$.
All fields are assumed to ``live'' on the light-ray, $\phi(z_i) \equiv \phi(nz_i)$.
The color tensors are defined as $C^{abc}_+ = if^{abc}$ and $C^{abc}_- = d^{abc}$.
The vector $s_T^\sigma = \varepsilon^{\sigma\mu\nu\lambda}s_\mu n_\nu \bar n_\lambda$\footnote{The following convention is used: $\varepsilon^{0123}=-1$.} is a projection of the nucleon spin $s$ ($s^2=-1$) transverse to the light-cone spanned by $n$ and a second light-like vector $\bar n$.

For the purpose of this work it is convenient to use $C$-even and $C$-odd operators:
\begin{align}
 \mathbb S^\pm_q(z) &= i s^\sigma_T \bigl(S^+_{q,\sigma}(z) \pm  P_{13}S^-_{q,\sigma}(z) \bigr)\,,\notag\\
 \mathbb F^\pm(z) &= (1 \mp P_{23} \pm P_{12})F^\pm(z)\,,
\end{align}
where $P_{ij}$ denotes the permutation operator acting on the position $z_i$ and $z_j$.
The matrix elements between two nucleon states define the twist-three correlation functions $\mathfrak{S}^\pm_q $ and $\mathcal{F}^\pm$ in momentum fraction space:
\begin{align}
 \bra{p,s_T} \mathbb S^\pm_q(z)\ket{p,s_T} &= 2p_+^2\int\mathcal{D}x\,e^{-ip_+\sum_kx_kz_k} \mathfrak{S}^\pm_q(x)\,,\notag\\
 \bra{p,s_T} \mathbb F^\pm_q(z)\ket{p,s_T} &= 2p_+^3\int\mathcal{D}x\,e^{-ip_+\sum_kx_kz_k} \mathcal{F}^\pm(x)\,,
\end{align}
where the integration measure $\mathcal{D}x$ is given by
\begin{align}
 \mathcal{D}x &= dx_1dx_2dx_3\,\delta(x_1+x_2+x_3)\,.\label{eq:intmeasure}
\end{align}
The definitions of $\mathfrak{S}^\pm_q (x)\equiv \mathfrak{S}^\pm_q (x_1,x_2,x_3)$ and $\mathcal{F}^\pm(x)\equiv \mathcal{F}^\pm(x_1,x_2,x_3)$ coincide with those of Ref.~\cite{Braun:2009mi}.

In the following we will formulate the problem in terms of $\mathfrak{S}^\pm_q$ and $\mathcal{F}^\pm$, since the mixing  under renormalization between the $\mathfrak S$-type and $\mathcal F$-type functions takes a rather compact form.
Generally the notation of~\cite{Braun:2009mi} is used and the conversion to other notations can be achieved by taking an appropriate linear combination.
From now on it is tacitly assumed that all functions correspond to the matrix elements of operators between two \emph{proton} states.

$\mathfrak{S}^\pm_q$ and $\mathcal{F}^\pm$ are real and effectively functions of two momentum fractions, see Eq.~\eqref{eq:intmeasure}, and the scale $\mu^2$.
They obey coupled sets of evolution equations
\begin{align}
 \frac{\partial}{\partial\ln\mu^2}\begin{pmatrix}
                                   \mathfrak{S}^\pm\\
                                   \mathcal{F}^\pm
                                  \end{pmatrix}
&= 
-\frac{\alpha_s(\mu^2)}{2\pi}\begin{pmatrix}
                              \mathbb{H}_{QQ}^\pm & \mathbb{H}_{QF}^\pm\\
                              \mathbb{H}_{FQ}^\pm & \mathbb{H}_{FF}^\pm
                             \end{pmatrix}
\begin{pmatrix}
                                   \mathfrak{S}^\pm\\
                                   \mathcal{F}^\pm
                                  \end{pmatrix}\,.\label{eq:evolutioneq}
\end{align}
Note that Eq.~\eqref{eq:evolutioneq} is written for the flavor singlet distribution $\mathfrak{S}^\pm = \mathfrak{S}^\pm_u+\mathfrak{S}^\pm_d$.
The corresponding flavor-nonsinglet distribution obeys a simpler equation, which decouples from the gluonic distributions.
Without loss of generality we assume the evolution to have the structure of Eq.~\eqref{eq:evolutioneq}.
The program will evolve both flavor sectors separately to disentangle up- and down-quarks.
The only quantity on the rhs.\ of Eq.~\eqref{eq:evolutioneq} that depends explicitly on $\mu^2$ is the running coupling $\alpha_s$.
At leading order it is given by
\begin{align}
 \alpha_s(\mu^2) &= \frac{4\pi}{b_0\ln(\mu^2/\Lambda_{\text{QCD}}^2)}\,,\label{eq:runningalphas}
\end{align}
where $b_0 = 11/3\,N_c-2/3\, n_f$ for QCD with $N_c$ colors and $n_f$ flavors. $\Lambda_{\text{QCD}}$ is the dimensional transmutation scale.
By changing variables from $\mu^2$ to the dimensionless \emph{evolution time} $t$,
\begin{align}
 t &= -\frac{2}{b_0}\ln\biggl(\frac{\alpha_s(\mu^2)}{\alpha_s(\mu_0^2)}\biggr)\label{eq:evoltime}\,,
\end{align}
where $\mu_0$ is some reference scale, one can recast Eq.~\eqref{eq:evolutioneq} into a more convenient form,
\begin{align}
 \biggl(\frac{\partial}{\partial t}+\mathbb{H}^\pm\biggr) \begin{pmatrix}
                                   \mathfrak{S}^\pm\\
                                   \mathcal{F}^\pm
                                  \end{pmatrix}
                                  &= 0\,, \label{eq:evolutioneq2}
\end{align}
where
\begin{align}
 \mathbb{H}^\pm &= \begin{pmatrix}
                              \mathbb{H}_{QQ}^\pm & \mathbb{H}_{QF}^\pm\\
                              \mathbb{H}_{FQ}^\pm & \mathbb{H}_{FF}^\pm
                             \end{pmatrix}\,.
\end{align}
For illustrative purposes we restrict ourselves to $\alpha_s$ at leading order.

Suppose at some initial time $t_0=0$ (corresponding to the initial scale $\mu_0$) one specifies the functions $\mathfrak{S}^\pm$ and $\mathcal{F}^\pm$, denoted by $\mathfrak{S}^\pm_0 = \mathfrak{S}^\pm(t=0),\mathcal{F}^\pm_0 = \mathcal{F}^\pm(t=0)$.
Then the formal solution of Eq.~\eqref{eq:evolutioneq2} will be given by exponentiation of the evolution kernels,
\begin{align}
 \begin{pmatrix}
  \mathfrak{S}^\pm(t)\\
  \mathcal{F}^\pm(t)
 \end{pmatrix}
 &=
 \exp\bigl(-t\,\mathbb{H}^\pm\bigr)\begin{pmatrix}
  \mathfrak{S}^\pm_0\\
  \mathcal{F}^\pm_0
 \end{pmatrix}\,.\label{eq:solution}
\end{align}
This expression will be the starting point for the numerical treatment.
Exact approaches like diagonalizing $\mathbb{H}^\pm$ will not be pursued here.
At best, methods analogous to~\cite{Derkachov:1999ze} may only be applicable in certain limits, e.g. at large-$N_c$.

\section{Method of solution\label{sec:method}}

For small deviations of $t$ from the initial $t_0$ the series expansion of the solution in Eq.~\eqref{eq:solution} can be reliably truncated at some finite order.
The remaining ingredient and most expensive operation is the action of a generic ``Hamiltonian'' $\mathbb{H}$.
We give a brief and schematic description of its evaluation below.

Due to translational invariance the momentum fraction arguments $(x_1,x_2,x_3)$ of the twist-three distributions $\mathfrak{S}^\pm,\mathcal{F}^\pm$ are subject to the constraint
\begin{align}
 x_1+x_2+x_3 &= 0\,.
\end{align}
For definiteness, we will eliminate $x_3$ in favor of $x_1,x_2$ and work with a compact notation:
\begin{align}
 \mathfrak{S}^\pm (x_1,x_2) &\equiv \mathfrak{S}^\pm(x_1,x_2,-x_1-x_2)\,, \label{eq:fx1x2}
\end{align}
and similar for $\mathcal{F}^\pm$.

In general $\mathbb{H}$ is a linear one-dimensional integral operator whose action will be computed numerically.
This is implemented by a discretization of the momentum fraction support.
The choice here will be an equally spaced $N_p\times N_p$ grid, which contains points
\begin{align}
 \bigl(x_{1,2}\bigr)_i &= \frac{2(i+1)}{N_p-1} + \frac{N_p+1}{1-N_p}\,,
\end{align}
where $ i\in\{0,\dots,N_p-1\}$.
Note that the maximum  and minimum values for $i$ correspond to the boundaries $\pm1$ of the support of the correlation functions.
Then, one can evaluate the action of $\mathbb H$ on a function $f$ (in very schematic notation)
\begin{align}
 \mathbb{H}f\bigl(x_i,x_j\bigr) &= \sum_k  w_k\mathbb{K}\bigl(x_i,x_j,x_k\bigr) f\bigl(x_k,x_i+x_j-x_k\bigr)\,,\label{eq:generich}
\end{align}
where $\mathbb{K}$ is the associated integral kernel of $\mathbb{H}$ and $w_k$ are certain weights for the sampling points of the integration, e.g.\ according to Simpson's rule or other Newton-Cotes formulas~\cite{abramowitzstegun}.
The structure appearing in Eq.~\eqref{eq:generich} follows from momentum conservation and the actual support of the integration/summation is determined by the support of $f$ as well as by the form of $\mathbb K$.
We refer to the appendix of~\cite{Braun:2009mi} for all possible kernels.

There is one subtlety specific to this type of evolution equations.
Certain values of momentum fractions, like the zero gluon momentum limit, usually emerging from ``plus-like'' prescriptions, require some caution, see~\cite{Braun:2009mi,Schafer:2012ra,Ma:2012xn,Kang:2012em}.
A possibility to deal with this would be to introduce some regulator, which avoids divisions by zero and ensures that all limits are taken correctly (up to errors of the order of the regulator).
The kernels implemented in this program treat this issue differently: they react on these ``special'' points appropriately with an analytic expression that needs to be hard-coded.

\section{Description of the program\label{sec:description}}

Usually the user wants to customize the program for his or her needs.
In order to do so, there are only a few places where one has to dig into the actual code, while one can widely ignore the low-level routines.
In the following an overview over the different parts of the program is presented and typical user-relevant modifications are pointed out.

\subsection{Source files}

The source of the program consists of $13$ files, which are listed below along with a short description of their contents.
\begin{verbatim}
 D1Function.cpp, D1Function.h
\end{verbatim}
contain a wrapper class for functions of one variable along with some useful routines (like calculating derivatives or saving to a file).
\begin{verbatim}
 HexFunction.cpp, HexFunction.h
\end{verbatim}
contain a wrapper class for functions of two variables with support on a hexagonal simplex. Examples are $\mathfrak{S}^\pm_q$, $\mathcal{F}^\pm$ and functions derived from it, see Sec.~\ref{sec:output}.
\begin{verbatim}
 qfqkernels.h, mixkernels.h,
 fffkernels.h, g2kernels.h
\end{verbatim}
implement the integral operators $\mathbb{H}^\pm_{QQ}$, $\mathbb{H}^\pm_{FQ,QF}$, $\mathbb{H}^\pm_{FF}$ and the integrals necessary for the extraction of $g_2$.
The initial conditions for the evolution can be given in
\begin{verbatim}
 initial.h,
\end{verbatim}
see Sec.~\ref{sec:initial} for details.
\begin{verbatim}
 constants.cpp, constants.h
\end{verbatim}
specify several global parameters for the evolution, see Sec.~\ref{sec:basicparameters}.
\begin{verbatim}
 mathutil.h
\end{verbatim}
contains the implementation of the numerical integration, viz. Eq.~\eqref{eq:generich}.
\begin{verbatim}
 t3evol.cpp
\end{verbatim}
represents the main program.

\subsection{How to compile}

In principle {\tt t3evol} can be compiled with any C++ compiler, for definiteness we assume it to be {\tt g++}.
The basic command to compile the program is
\begin{verbatim}
 g++ t3evol.cpp HexFunction.cpp 
 D1Function.cpp constants.cpp
\end{verbatim}
which does not produce the fastest code.
The optimization options 
\begin{verbatim}
 -O3 -fno-trapping-math 
 -fomit-frame-pointer -funroll-loops
\end{verbatim}
provided by {\tt g++} can be used safely and produced an executable that did the calculation twice as fast compared to the unoptimized case.
A further performance enhancement (up to a few per cent) can be achieved by using profiles.
It is recommended to use all of the above.

In addition the unsafe option 
\begin{verbatim}
 -ffast-math
\end{verbatim}
produced the fastest code in the test case.
The result of the computation did not change compared to the ``safe'' option scenario (although it could have).
There is no warranty that this may be true in a general situation, the use of this option is therefore not recommended. 

\subsection{Basic parameters\label{sec:basicparameters}}

The files {\tt constants.h} and {\tt constants.cpp} contain some basic parameters for the simulation, which are listed and described below.
\begin{verbatim}
 Np 
\end{verbatim}
is the number of discretization points.
Is is identical to $N_p$ from Sec.~\ref{sec:method}.
It should be an odd number, in order to have the central point $(x_1,x_2)=(0,0)$ and the boundary points $(\pm1,.)$, $(.,\pm1)$ lying on the grid.
The central point is characterized by an integer {\verb ZEROPOS }, which is equal to $(N_p-1)/2$ and should not be changed.
\begin{verbatim}
 NumSteps
\end{verbatim}
specifies the number of terms in the series of the formal solution in Eq.~\eqref{eq:solution}. For example, if {\verb NumSteps } is $4$ the exponential will be computed to the accuracy \mbox{$\mathcal{O}\bigl((t-t_0)^4\bigr)$}, neglecting terms of order $\mathcal{O}\bigl((t-t_0)^5\bigr)$.
\begin{verbatim}
 mN
\end{verbatim}
is the nucleon mass $m_N$ in units of $\giga\electronvolt$.
Its default value is set to $\unit{0.938}{\giga\electronvolt}$.
\begin{verbatim}
 Nc, oneoverNc, Cf
\end{verbatim}
are parameters of the special unitary group $\mathit{SU}(N_c)$, namely $N_c$, $1/N_c$ and $C_F=(N_c^2-1)/(2N_c)$, respectively.
Formally a large $N_c$ limit can be taken by setting {\verb oneoverNc } to zero and {\verb Cf } to $N_c/2$.
Of course, the default value is $N_c=3$.
\begin{verbatim}
 Nf
\end{verbatim}
is the number of different flavors participating in the evolution.
Since only up- and down-quark correlations are supported, it should be set to $2.0$.
It is treated as an independent parameter and does not interfere with the number of flavors in $\alpha_s(\mu^2)$, see below.

For the running coupling $\alpha_s(\mu^2)$ a couple of implementations at leading order and next-to-leading order are available.
The default treatment is analogous to GRV98LO, cf.~\cite{Gluck:1998xa} for details and values of the quantities below.
At lowest order the beta function is given in
\begin{verbatim}
 double beta0(double nf),
\end{verbatim}
which depends on the number of active flavors at the scale $\mu$, which in turn is given by
\begin{verbatim}
 int nf(double mu),
\end{verbatim}
which changes its value at the flavor thresholds
\begin{verbatim}
 FTH34, FTH45, FTH56.
\end{verbatim}
The matching parameters are adjusted accordingly in
\begin{verbatim}
 double lambda_LO_GRV98(int nf).
\end{verbatim}
From these parameters the running coupling and the evolution time is computed in 
\begin{verbatim}
 double alpha_s_LO_GRV98(double mu)
 double evolutionTime_LO_GRV98(double
 mu0, double mu1, int nf),
\end{verbatim}
\noindent
viz. Eqs.~\eqref{eq:runningalphas}, \eqref{eq:evoltime}.
Other implementations can also be used by replacing each call of the latter function in the main program.

The rest of the constants are used for the numerical integration, cf.\ $w_k$ in Eq.~\eqref{eq:generich}, and can be ignored by the user.

\subsection{Initial distributions\label{sec:initial}}

There are two ways to specify the initial conditions, $\mathfrak{S}^\pm_0 $, $\mathcal{F}^\pm_0$.
The first way is to implement the functions inside the file {\tt initial.h}, namely
\begin{verbatim}
 initialFunction_Splus_u
 initialFunction_Splus_d
 initialFunction_Sminus_u
 initialFunction_Sminus_d
 initialFunction_Fplus
 initialFunction_Fminus
\end{verbatim}
for  $\mathfrak{S}^+_u(t_0)$, $\mathfrak{S}^+_d(t_0)$, $\mathfrak{S}^-_u(t_0)$, $\mathfrak{S}^-_d(t_0)$, $\mathcal{F}^+(t_0)$, $\mathcal{F}^-(t_0)$ respectively.
All of them are real functions of $x_1,x_2,x_3$.
The prescription of Eq.~\eqref{eq:fx1x2} is done by function overloading inside the same file.
By default the program is shipped with the model of~\cite{Braun:2011aw} and $\mathcal{F}^\pm(t_0) = 0$.

The second way is to read data from a file.
The program will check whether inside its directory one (or more) of the following files exist:
\begin{verbatim}
 S+_singlet_initial.txt
 S+_nonsinglet_initial.txt
 S-_singlet_initial.txt
 S-_nonsinglet_initial.txt
 F+_initial.txt
 F-_initial.txt
\end{verbatim}
They correspond to  $\mathfrak{S}^+_u(t_0)+\mathfrak{S}^+_d(t_0)$, $\mathfrak{S}^+_u(t_0)-\mathfrak{S}^+_d(t_0)$, $\mathfrak{S}^-_u(t_0)+\mathfrak{S}^-_d(t_0)$, $\mathfrak{S}^-_u(t_0)-\mathfrak{S}^-_d(t_0)$, $\mathcal{F}^+(t_0)$, $\mathcal{F}^-(t_0)$.
The scope of these input files should be such that it contains numbers in $N_p$ rows and $N_p$ columns, where the number at position $(i,j)$ corresponds to the value of the function at $\bigl((x_1)_i,(x_2)_j\bigr)$.
In fact, the program reads $N_p^2$ tokens, and any data beyond these will be ignored.

Note that the initialization files are automatically used if they are present.
If one does not want to use them, they need to be removed from the directory.
Therefore it is also possible to use a ``hybrid'' mode, giving some functions by file and the others by code.

At this point it is important to note that $\mathfrak{S}^\pm$ and $\mathcal{F}^\pm$ have to obey certain symmetry relations and restrictions on the support, cf.~\cite{Braun:2009mi}.
The user himself is responsible to ensure that these properties are fulfilled in his input.
Apart from that there is no restriction on the functional form for the initial conditions.
In that sense the evolution can be tested on any model, which seems to be necessary given the fact that the available models and estimates differ drastically~\cite{Braun:2011aw,Kanazawa:2010au,Boer:2011fx,Kang:2012xf,Metz:2012ui}.

\subsection{Running the program}

{\tt t3evol} can effectively be called with zero, one or two command line parameters.
If two parameters are given, it will assume that the first one is the initial scale $\mu_0^2$ (in $\giga\electronvolt^2$) and the second one is the final scale $\mu^2$ (in $\giga\electronvolt^2$).
If only one parameter is received, it is assumed that it corresponds to $\mu_0^2$ and the program awaits an input from the user for $\mu^2$.
When no parameter is specified, the program will ask for both $\mu_0^2$ and $\mu^2$.
Note that it is required that $\mu^2>\mu_0^2$ and obviously $\mu_0^2>0$.

The program then continues to do the evolution using the methods of Sec.~\ref{sec:method}, making a separate calculation for each flavor and each $C$-parity sector.
Solutions are written as soon as they are available.

\section{Output\label{sec:output}}

The results of the evolution will be written to several files in the directory of the program, each function to a separate file.
For convenience, this is done for a couple of equivalent representations owing to the variety of conventions in the literature.
One may safely modify the main program to restrict the output to the quantities of the user's interest,
The output filenames are of the form
\begin{verbatim}
 prefix_final.txt,
\end{verbatim}
where {\verb prefix } stands for an identifier for one of the functions widely used in the literature.
The multitude of functions is listed in Tab.~\ref{tab:prefixes}.
\begin{table}
\begin{center}
\begin{tabular}{ccc}
\hline\hline
 {\verb prefix }         & function                             & definition  \\
 \hline
  {\tt DeltaTdFd}        & $\Delta T_{\bar dFd}$                & \cite{Kang:2008ey,Braun:2009mi}\\
  {\tt DeltaTuFu}        & $\Delta T_{\bar uFu}$                & \cite{Kang:2008ey,Braun:2009mi}\\
  {\tt F-}               & $\mathcal{F}^-$                      & \cite{Braun:2009mi}\\
  {\tt F+}               & $\mathcal{F}^+$                      & \cite{Braun:2009mi}\\
  {\tt GFd}              & $G^d_{F}$                            & \cite{Kanazawa:2010au}\\
  {\tt GFu}              & $G^u_{F}$                            & \cite{Kanazawa:2010au}\\
  {\tt N}                & $N$                                  & \cite{Beppu:2010qn}\\
  {\tt O}                & $O$                                  & \cite{Beppu:2010qn}\\
  {\verb S-_d }          & $\mathfrak{S}^-_d$                   & \cite{Braun:2009mi}\\
  {\verb S+_d }          & $\mathfrak{S}^+_d$                   & \cite{Braun:2009mi}\\
  {\verb S-_nonsinglet } & $\mathfrak{S}^-_u -\mathfrak{S}^-_d$ & here\\
  {\verb S+_nonsinglet } & $\mathfrak{S}^+_u -\mathfrak{S}^+_d$ & here\\
  {\verb S-_singlet }    & $\mathfrak{S}^-_u +\mathfrak{S}^-_d$ & here\\
  {\verb S+_singlet }    & $\mathfrak{S}^+_u +\mathfrak{S}^+_d$ & here\\
  {\verb S-_u }          & $\mathfrak{S}^-_u$                   & \cite{Braun:2009mi}\\
  {\verb S+_u }          & $\mathfrak{S}^+_u$                   & \cite{Braun:2009mi}\\
  {\tt T3F- }            & $T_{3F}^-$                           & \cite{Braun:2009mi}\\
  {\tt T3F+ }            & $T_{3F}^+$                           & \cite{Braun:2009mi}\\
  {\tt TdFd}             & $T_{\bar dFd}$                       & \cite{Kang:2008ey,Braun:2009mi}\\
  {\tt TuFu}             & $T_{\bar uFu}$                       & \cite{Kang:2008ey,Braun:2009mi}\\
 \hline 
\end{tabular}
\caption{\label{tab:prefixes}Prefix identifier for the generated output functions and reference(s) to their definition.}
\end{center}
\end{table}
These files have a structure of 
\begin{align*}
 &x_1\quad x_2 \quad f(x_1,x_2)
\end{align*}
in each line, following the conventions of Eq.~\eqref{eq:fx1x2}.
In addition, for some of the antiquark-gluon-quark type correlations the so-called soft-gluon-pole and soft-fermion-pole configurations are exported in files of the form
\begin{verbatim}
 prefix_final_SGP.txt,
 prefix_final_SFP.txt.
\end{verbatim}
Their content is written in the form
\begin{align*}
 &x \quad f(x,0) \quad\text{(for {\tt SGP})}\,,\\
 &x \quad f(0,x) \quad\text{(for {\tt SFP})}
\end{align*}
in each line, again following the conventions of Eq.\ \eqref{eq:fx1x2}.

As a byproduct one can extract the twist-three contribution to the structure function $g_2$, see~\cite{Braun:2011aw}, via the relation
\begin{align}
 g_2^{\text{tw-3}}(x_B) &=  \sum_q \frac{e_q^2}{2m_N}\int \mathcal{D}x\,\mathfrak{S}_q^+(x)\notag\\
 &\quad \times\biggl(\frac{1 - P_{13}}{x_2} + \frac{d}{dx_3}\biggr)\frac{\theta(x_3 - x_B)}{x_2x_3}\,.
\end{align}
The resulting functions are written to
\begin{verbatim}
 xg2_tw3_proton_final.txt,
 xg2_tw3_neutron_final.txt,
\end{verbatim}
where $g_2$ for the neutron is obtained by a simple isospin rotation from the proton correlators.

\section{Conclusions\label{sec:conclusion}}

To the best of the author's knowledge, {\tt t3evol} represents the first closed solution to the twist-three renormalization group for the single spin asymmetry inducing multi-parton correlators.
It is a first step towards a correct incorporation of QCD evolution effects in phenomenological applications like a global fitting procedure.

The program, especially the implementation of the Hamiltonians, has been thoroughly tested for simple input functions, for which one can calculate the action of $\mathbb H$ analytically.
An example output for a realistic ansatz has already been presented in a model study based on light-cone wave functions~\cite{Braun:2011aw}.
In the particular case of $g_2$, a good agreement with effective approaches~\cite{Braun:2001qx} has been found.
The reproducibility of the anomalous dimensions of $g_2$ is a very strong check and indicates that the program is working correctly.
Further checks may be in order to clarify this claim.

The advantage to choose an arbitrary set of initial distributions is probably the biggest limitation on the performance.
If there was a more or less universal functional form for the correlators, it may be possible to invent more efficient approaches to deal with the evolution in terms of moments analogous to~\cite{Vogt:2004ns}.
Unfortunately the generalization of the Mellin-techniques to the problem at hand does not seem to be straightforward and requires further investigation.

An obvious way to increase the performance of the code in a multi processor environment is to use multithreading.
Since there are essentially four sectors (flavor (non-)singlet, $C$-parity $\pm$) that evolve independently, a parallelization of these calculations suggests itself.
This feature will be reserved for future versions.

\section*{Acknowledgements}

The author is grateful to V.M.~Braun and A.N.~Manashov for collaboration and to M.~Gruber for valuable discussions.
This project is partially supported by the DFG, grant BR2021/5-2 ``Multiparton evolution equations in QCD''.


\begin{thebibliography}{99}


\bibitem{Cafarella:2003jr}
  A.~Cafarella and C.~Coriano,
  Comput.\ Phys.\ Commun.\  {\bf 160}, 213 (2004).

\bibitem{Vogt:2004ns}
  A.~Vogt,
  Comput.\ Phys.\ Commun.\  {\bf 170}, 65 (2005).

\bibitem{Salam:2008qg} 
  G.~P.~Salam and J.~Rojo,
  Comput.\ Phys.\ Commun.\  {\bf 180}, 120 (2009).
  
\bibitem{Botje:2010ay} 
  M.~Botje,
  Comput.\ Phys.\ Commun.\  {\bf 182}, 490 (2011).

\bibitem{Liang:2000gz} 
  Z.~-T.~Liang and C.~Boros,
  Int.\ J.\ Mod.\ Phys.\ A {\bf 15}, 927 (2000)
  
\bibitem{Anselmino:1994gn} 
  M.~Anselmino, A.~Efremov and E.~Leader,
  Phys.\ Rept.\  {\bf 261}, 1 (1995)
  [Erratum-ibid.\  {\bf 281}, 399 (1997)]

\bibitem{Barone:2001sp} 
  V.~Barone, A.~Drago and P.~G.~Ratcliffe,
  Phys.\ Rept.\  {\bf 359}, 1 (2002)

\bibitem{Efremov:1981sh} 
  A.~V.~Efremov and O.~V.~Teryaev,
  Sov.\ J.\ Nucl.\ Phys.\  {\bf 36}, 140 (1982)
  [Yad.\ Fiz.\  {\bf 36}, 242 (1982)].


\bibitem{Qiu:1991wg} 
  J.~-W.~Qiu and G.~F.~Sterman,
  Nucl.\ Phys.\ B {\bf 378}, 52 (1992).

\bibitem{Bukhvostov:1984as} 
  A.~P.~Bukhvostov, E.~A.~Kuraev and L.~N.~Lipatov,
  JETP Lett.\  {\bf 37}, 482 (1983)
  [Pisma Zh.\ Eksp.\ Teor.\ Fiz.\  {\bf 37}, 406 (1983)]
  [Sov.\ Phys.\ JETP {\bf 60}, 22 (1984)]
  [Zh.\ Eksp.\ Teor.\ Fiz.\  {\bf 87}, 37 (1984)].

\bibitem{Bukhvostov:1985rn} 
  A.~P.~Bukhvostov, G.~V.~Frolov, L.~N.~Lipatov and E.~A.~Kuraev,
  Nucl.\ Phys.\ B {\bf 258}, 601 (1985).

  
\bibitem{Balitsky:1987bk} 
  I.~I.~Balitsky and V.~M.~Braun,
  Nucl.\ Phys.\ B {\bf 311}, 541 (1989).
  
\bibitem{Braun:2009mi} 
  V.~M.~Braun, A.~N.~Manashov and B.~Pirnay,
  Phys.\ Rev.\ D {\bf 80}, 114002 (2009)
  [Erratum-ibid.\ D {\bf 86}, 119902 (2012)]
  
\bibitem{Ma:2012xn} 
  J.~P.~Ma and Q.~Wang,
  Phys.\ Lett.\ B {\bf 715}, 157 (2012)

\bibitem{Kang:2008ey} 
  Z.~-B.~Kang and J.~-W.~Qiu,
  Phys.\ Rev.\ D {\bf 79}, 016003 (2009)
  
\bibitem{Vogelsang:2009pj} 
  W.~Vogelsang and F.~Yuan,
  Phys.\ Rev.\ D {\bf 79}, 094010 (2009)


\bibitem{Schafer:2012ra} 
  A.~Sch\"afer and J.~Zhou,
  Phys.\ Rev.\ D {\bf 85}, 117501 (2012)

\bibitem{Kang:2012em} 
  Z.~-B.~Kang and J.~-W.~Qiu,
  Phys.\ Lett.\ B {\bf 713}, 273 (2012)

\bibitem{Braun:2011aw} 
  V.~M.~Braun, T.~Lautenschlager, A.~N.~Manashov and B.~Pirnay,
  Phys.\ Rev.\ D {\bf 83}, 094023 (2011)

\bibitem{Derkachov:1999ze} 
  S.~E.~Derkachov, G.~P.~Korchemsky and A.~N.~Manashov,
  Nucl.\ Phys.\ B {\bf 566}, 203 (2000)
  
\bibitem{abramowitzstegun}
 M.~Abramowitz  and I.~A.~Stegun,
 Handbook of Mathematical Functions (1972)

\bibitem{Gluck:1998xa}
  M.~Gl\"uck, E.~Reya and A.~Vogt,
  Eur.\ Phys.\ J.\  C {\bf 5}, 461 (1998)

\bibitem{Kanazawa:2010au} 
  K.~Kanazawa and Y.~Koike,
  Phys.\ Rev.\ D {\bf 82}, 034009 (2010)
  
\bibitem{Boer:2011fx} 
  D.~Boer,
  Phys.\ Lett.\ B {\bf 702}, 242 (2011)
  
\bibitem{Kang:2012xf} 
  Z.~-B.~Kang, A.~Prokudin and ,
  Phys.\ Rev.\ D {\bf 85}, 074008 (2012)
  
\bibitem{Metz:2012ui} 
  A.~Metz, D.~Pitonyak, A.~Schafer, M.~Schlegel, W.~Vogelsang and J.~Zhou,
  Phys.\ Rev.\ D {\bf 86}, 094039 (2012)

\bibitem{Beppu:2010qn} 
  H.~Beppu, Y.~Koike, K.~Tanaka and S.~Yoshida,
  Phys.\ Rev.\ D {\bf 82}, 054005 (2010)

\bibitem{Braun:2001qx} 
  V.~M.~Braun, G.~P.~Korchemsky and A.~N.~Manashov,
  Nucl.\ Phys.\ B {\bf 603}, 69 (2001)

\end{thebibliography}
\end{document}